\documentclass[preprint,12pt]{elsarticle}

\usepackage[T1]{fontenc}
\usepackage[utf8]{inputenc}

\usepackage{amssymb}
\usepackage{amsmath}
\usepackage{cancel}
\usepackage{notoccite}
\usepackage{xcolor}
\usepackage[final]{changes}
\definechangesauthor[name={Reviewer 1}, color=red]{R1}
\definechangesauthor[name={Reviewer 2}, color=blue]{R2}

\setcitestyle{square}
\journal{JCP}

\begin{document}

\begin{frontmatter}



\title{Galilean Electromagnetic Particle-in-Cell Code} 


\author{Alexander Pukhov, Nina Elkina and Thomas Wilson} 

\affiliation{organization={Institute for Theoretical Physics I, University of D\"usseldorf},
            city={D\"usseldorf},
            postcode={40225}, 
            country={Germany}}

\begin{abstract}
We introduce a Galilean electromagnetic particle-in-cell (GEM-PIC) algorithm, which transforms the full set of Maxwell equations and the Vlasov equation into the boosted coordinates. This approach preserves the electromagnetic structure of the interaction while exploiting scale separation for computational efficiency. Unlike quasistatic methods, GEM-PIC does not have to distinguish between “beam” and “streaming” particles, allowing a self-consistent treatment of particle trapping. The GEM-PIC algorithm allows for highly efficient and accurate simulations of plasma-based wakefield acceleration.
\end{abstract}




\end{frontmatter}



\section{Introduction}

Plasma-based particle acceleration is a rapidly developing area of
modern science \cite{Joshi:2019}. Unlike conventional solid-state accelerators,
plasmas can support electric fields several orders of magnitude stronger.
This opens the door to compact, high-energy accelerators with significantly
reduced size and cost. Plasma wakefields can be excited either by
a high-current charged particle beam \textemdash{} beam-driven plasma
wakefield acceleration (PWFA)~\cite{Katsouleas:1987} 
\textemdash{} or by a high-intensity
laser pulse \textemdash{} laser wakefield acceleration (LWFA)~\cite{Tajima:1979}. The
current record for LWFA is a 10\,GeV energy gain achieved in a mere
20\,cm-long plasma channel \cite{Bella:2023, Rockafellow:2025, Korea:2019}.
In contrast, reaching such energies in a conventional radio-frequency
(RF) accelerator would require several kilometers of infrastructure.
In PWFA, the landmark Stanford experiment demonstrated a doubling
of a 42\,GeV electron bunch energy in just a 1-meter plasma cell 
\cite{Blumenfeld:2007}. 
The new AWAKE project at CERN utilizes proton bunch self-modulation to accelerate leptons and aims to achieve electron energies in the TeV range \cite{AWAKE_Symmetry:2022}.



The most widely used simulation tools for plasma wakefield acceleration
are full electromagnetic particle-in-cell (PIC) codes 
Osiris \cite{Osiris}, VLPL~\cite{VLPL},
Warp-X~\cite{WarpX}, SMILEI~\cite{SMILEI}, FBPIC~\cite{FBPIC}. These solve Maxwell\textquoteright s equations
on a spatial grid while tracking the relativistic motion of macroparticles
that represent the plasma. As \textit{ab initio} models, PIC codes can
include the complete physics of the interaction. However, this fidelity
comes at a high computational cost \textemdash{} so high that realistic
LWFA simulations often require exascale supercomputers.

This computational challenge arises from the intrinsic multiscale
nature of the problem. The typical acceleration length is $L_{\text{acc}}\approx10\,\mathrm{cm}$,
the plasma wavelength is $\lambda_{p}\approx30\,\mu\mathrm{m}$, and
the laser wavelength is $\lambda_{L}\approx1\,\mu\mathrm{m}$. A fully
electromagnetic PIC code must resolve both $\lambda_{L}$ and the
corresponding laser period $\tau_{L}=\lambda_{L}/c$, where $c$ is
the speed of light. With a conservative estimate of 10 steps per laser
period, simulating the full acceleration length requires $N\sim10^{6}$
time steps. Considering that a typical simulation tracks $\sim10^{9}$
particles, and each particle update requires $\sim10^{3}$ floating-point
operations per step, the total computational cost is approximately
$10^{18}$ FLOPs \textemdash{} placing the task firmly in the exascale
regime.

Several strategies have been developed to mitigate this cost. FBPIC,
for example, uses a cylindrical geometry with a limited number of
azimuthal modes, significantly improving performance. SMILEI and similar
codes employ the envelope approximation for the laser pulse, enabling
coarser grids and larger time steps. Warp-X introduced a Lorentz-boosted
frame approach in which the simulation is performed in a relativistically
moving frame. This exploits Lorentz contraction to reduce the simulated
length scale. However, this method introduces its own challenges:
the background plasma becomes highly energetic and streamed 
relativistically, increasing numerical noise. 
Moreover, transforming data
back to the laboratory frame is non-trivial due to the relativity of 
simultaneity \cite{Yu:2015,Lehe:2016}.

A major breakthrough in simulating long-distance laser and beam propagation
in tenuous plasma occurred with the development of the first quasi-static
PIC code, WAKE \cite{Wake}. This method introduced a Galilean coordinate
transformation using a fast coordinate $\zeta=z-ct$ and a slow time
$\tau=t$. By assuming a static plasma response at each step in slow
time, the method effectively decouples fast and slow dynamics,
drastically reducing computational costs. However, the quasi-static
approximation comes with limitations. The laser pulse must be modeled
via the envelope approximation, preventing accurate modeling of radiation
emission. More generally, all quasi-static codes\textemdash such as
QuickPIC~\cite{QuickPIC}, HiPACE~\cite{HiPACE}, Wake-T~\cite{WakeT}, 
LCODE~\cite{LCODE3D}, and QV3D~\cite{Pukhov:2016} assume a static
response to a fixed driver (either a particle bunch or a laser envelope).
These methods separate particles into \textquotedblleft streaming\textquotedblright{}
background plasma and \textquotedblleft beam\textquotedblright{} driver
particles. Streaming particles are updated along the fast coordinate
$\zeta$, while beam particles evolve in slow time. This separation
breaks down in situations involving self-consistent particle trapping
or radiation, since the critical terms in the Maxwell equations
are omitted.

A first attempt to incorporate radiation into a quasi-static framework
was made in \cite{Tuev:2023}. However, the authors still relied on separating
streaming and beam particles, thus excluding self-consistent trapping.
Furthermore, the implementation was limited to two-dimensional (2D)
geometries\textemdash either Cartesian $(X,Z)$ or cylindrical $(r,Z)$.

In this work, we propose a fundamentally new approach. We introduce
a Galilean electromagnetic particle-in-cell (GEM-PIC) algorithm, which
transforms the full set of Maxwell equations and the Vlasov equation
into the boosted coordinates.  This approach preserves the full electromagnetic
structure of the interaction while exploiting scale separation
for computational efficiency. Unlike quasistatic methods, GEM-PIC
does not have to distinguish between \textquotedblleft beam\textquotedblright{}
and \textquotedblleft streaming\textquotedblright{} particles, enabling
a self-consistent treatment of particle trapping. The algorithm supports
flexible step sizes along both the fast and slow coordinates. The
step along the slow coordinate can be governed by physical timescales
such as the betatron period or the Rayleigh length of the laser pulse
and is typically orders of magnitude larger than that in conventional
PIC codes. The step along the fast coordinate can be locally refined
to resolve short wavelengths \textemdash{} including x-rays \textemdash{}
 potentially enabling efficient simulations of systems such as plasma-based XFELs.

\section{Properties of Gallilean transformation}

When we simulate a laser pulse or a bunch of particles propagating
through low-density plasma, they evolve rather slowly. The characteristic
evolution time for an electron bunch of particles is defined by the
betatron frequency $\omega_{\beta}=\omega_{p}/2\gamma^{1/2}$, where
$\omega_{p}=\sqrt{4\pi ne^{2}/m}$ is the plasma frequency for the plasma
with background electron density $n$, and $\gamma$ is the relativistic
gamma-factor for electrons of the bunch. For highly relativistic particles,
$\gamma\gg1$, we have $\omega_{\beta}\ll\omega_{p}$. Thus, a significant
evolution of the bunch occurs over many plasma wake periods. If we
take a laser pulse in a very underdense plasma, $\omega_{p}\ll\omega_{L}$,
the characteristic evolution distance for the laser pulse is the Rayleigh
length $R_{0}=\pi\sigma_{\perp}^{2}/\lambda_{L}$. Here, $\sigma_{\perp}$
is the laser focal spot radius.

If we assume that the laser pulse, the driver particle bunch, and
the witness bunch all propagate in the same direction along the $Z-$axis
at relativistic velocities, it is natural to introduce a set of comoving
coordinates. In this framework, a \textquotedblleft fast\textquotedblright{} coordinate $\zeta$  captures the internal structure of the driver and the plasma perturbations it induces, while a \textquotedblleft slow\textquotedblright{} coordinate $s$  describes the long-scale evolution of these structures over time or distance.

Two principal Galilean coordinate transformations can be considered.
The first "spatial" one is:

\begin{align}
  s &= ct\\
 \zeta &= z-ct  \label{eq:Galilean1}
 \end{align}
 and in matrix form:
 \begin{align}
\text{GT1:}&\quad 
\begin{bmatrix}
s\\
\zeta
\end{bmatrix} = 
\begin{bmatrix}
c &0\\
-c &1
\end{bmatrix}
\begin{bmatrix}
t\\
z
\end{bmatrix}\\
\text{inverse GT1:}&\quad 
\begin{bmatrix}
t\\
z
\end{bmatrix} = 
\begin{bmatrix}
1/c &0\\
1 &1
\end{bmatrix}
\begin{bmatrix}
s\\
\zeta
\end{bmatrix}
\end{align}
The derivatives for GT1 are transformed as
\footnote{
\begin{align*}
s &=ct,\quad &\frac{\partial}{\partial s} &= 
\frac{\partial t}{\partial s}\frac{\partial }{\partial t} + 
\frac{\partial z}{\partial s}\frac{\partial }{\partial z} =
\frac{1}{c}\frac{\partial}{\partial t} + \frac{\partial }{\partial z}\\
\zeta &=z-ct\quad &\frac{\partial}{\partial \zeta} &= 
\frac{\partial t}{\partial \zeta}\frac{\partial }{\partial t} +
\frac{\partial z}{\partial \zeta}\frac{\partial }{\partial z} = 
 \frac{\partial }{\partial z}
\end{align*}
}
\begin{align}
\frac{\partial}{\partial t} &= \frac{\partial }{\partial s} - c\frac{\partial}{\partial \zeta},\\
\frac{\partial }{\partial z} &= \frac{\partial}{\partial \zeta}
\end{align}

An alternative "temporal" Galilean transformation is given below
\begin{align}
 \tilde s &= z\\
\tilde \zeta &= z-ct.
\label{eq:Galilean2}
\end{align}
In  matrix form, this transformation is given by
\begin{align}
\text{GT2:}&\quad 
\begin{bmatrix}
\tilde s\\
\tilde \zeta
\end{bmatrix} = 
\begin{bmatrix}
0 &1\\
-c &1
\end{bmatrix}
\begin{bmatrix}
t\\
z
\end{bmatrix}\\
\text{inverse GT2:} &\quad 
\begin{bmatrix}
t\\
z
\end{bmatrix} = 
\begin{bmatrix}
1 &0\\
1/c &-1/c
\end{bmatrix}
\begin{bmatrix}
\tilde s\\
\tilde \zeta
\end{bmatrix}
\end{align}
And the derivatives for GT2 are transformed accordingly.
\begin{align}
\frac{\partial }{\partial t} &= -c\frac{\partial}{\partial \tilde \zeta},\\
\frac{\partial}{\partial z} &= \frac{\partial }{\partial \tilde s} + \frac{\partial }{\partial \tilde\zeta}.
\end{align}

The action of each Galilean transformation on a fundamental solution of the Maxwell equations
\begin{align}
F(t,z) = \underbrace{F(z-ct)}_{\text{forward}} + \underbrace{F(z+ct)}_{\text{backward}}
\end{align}
is described accordingly:
\begin{align}
\text{GT1:}\qquad F(s,\zeta) &= F(\zeta) + F(2s+\zeta)\\
\text{GT2:}\qquad F(\tilde s,\tilde\zeta) &= F(\tilde\zeta) + 
F(2\tilde s-\tilde\zeta).
\end{align}
In both cases, the forward wave depends only on the fast variable. If GT1 is applied, the backward wave moves  backward at $-2c$ along $\zeta$. 
In case of GT2, the backward wave moves forward. 
\begin{figure}[ht]
\includegraphics[width=5cm]{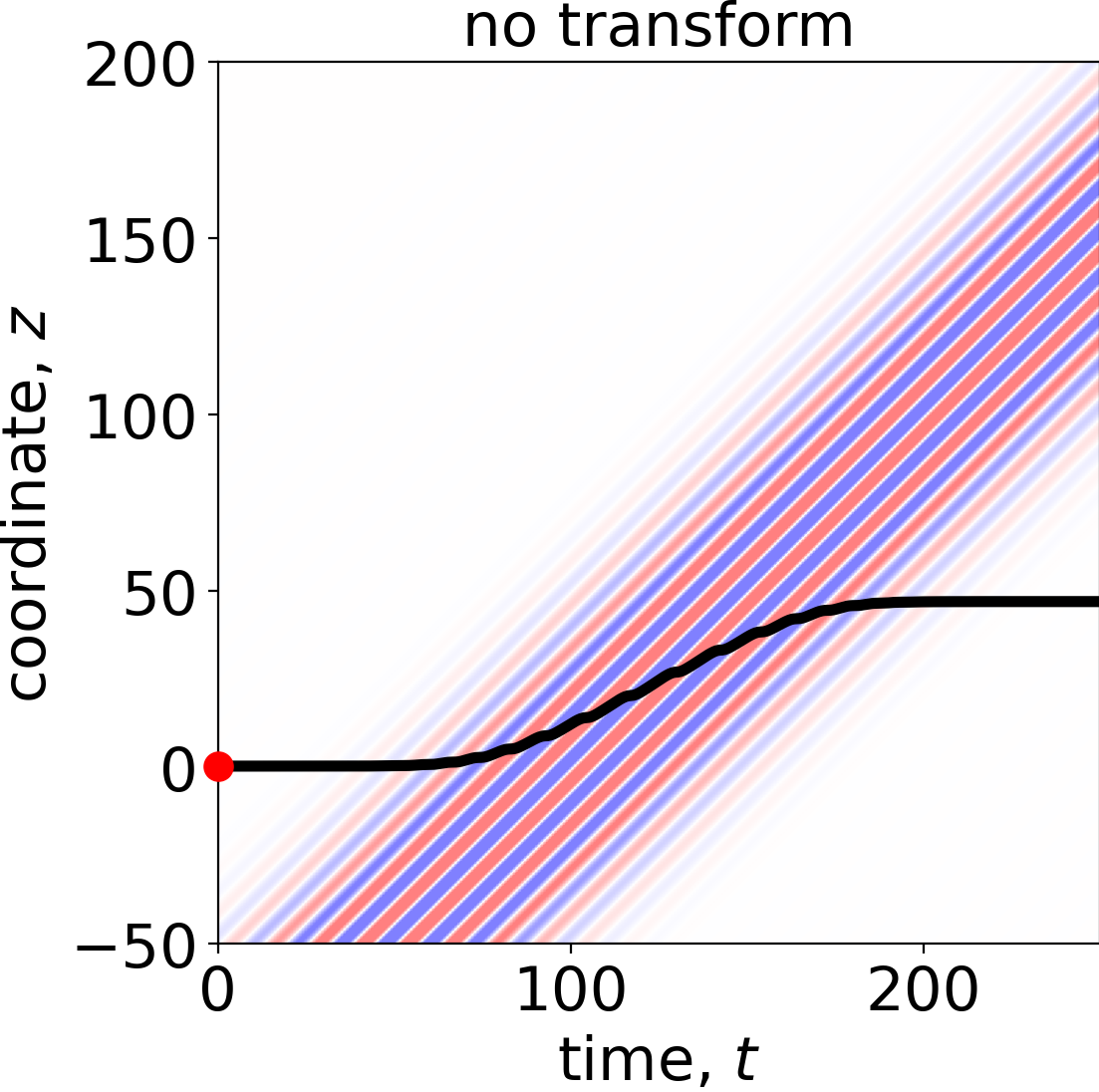}
\includegraphics[width=7.cm]{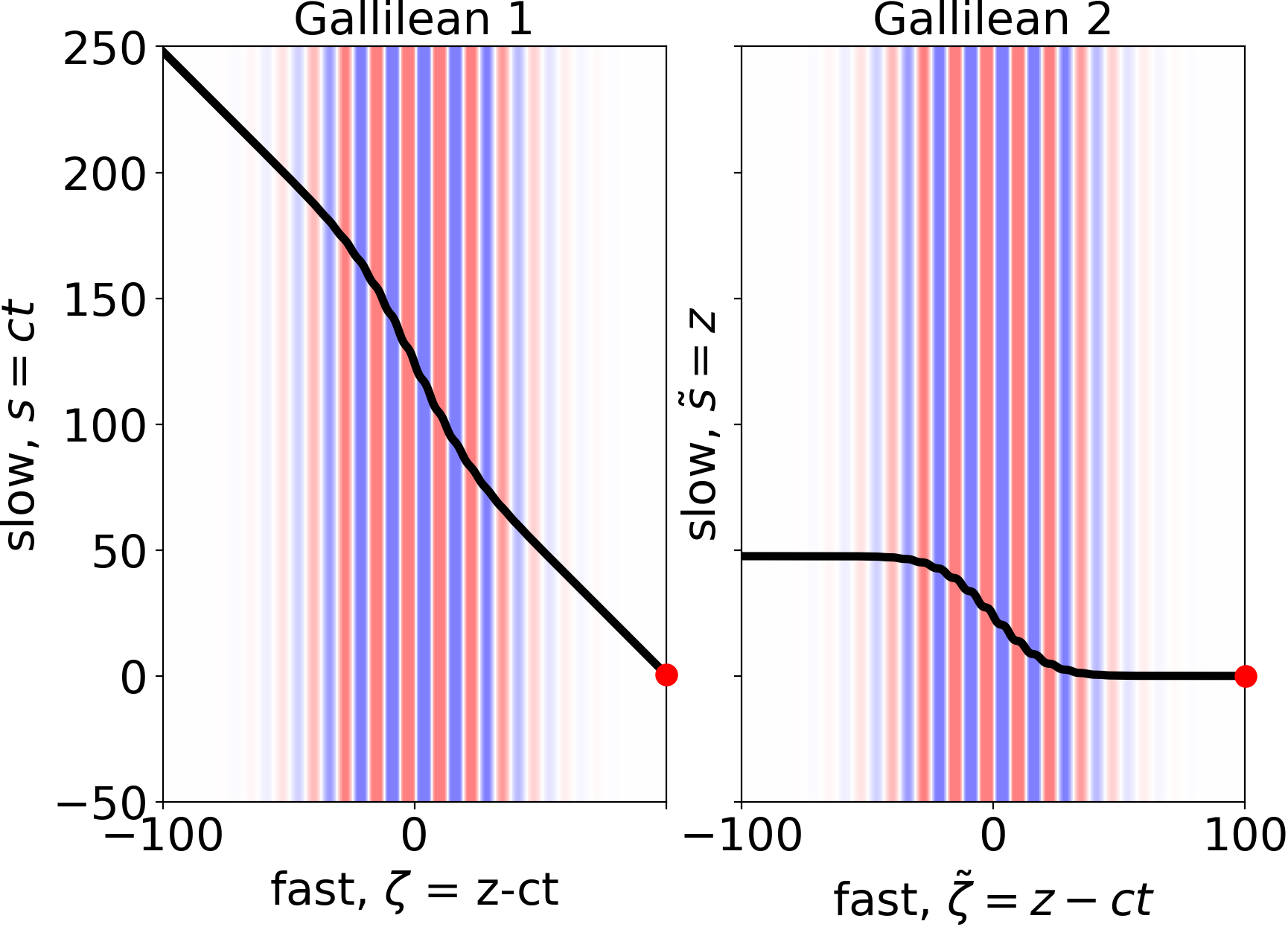}
\caption{The action of spatial-like GT1 (middle) and time-like GT2 (right) Galilean transformations on a particle and laser pulse compared with lab frame.}
\label{figGal}
\end{figure}
The space-time trajectory of a particle in a lab 
frame is given by 
\begin{eqnarray}
\frac{d\vec u}{dt} &=& \vec F \\
\frac{d x}{dt} &=& \frac{u_x}{\gamma} \\
\frac{d \vec r_\perp}{dt} &=& \frac{\vec u_\perp}{\gamma},
\end{eqnarray}
can be determined in the transformed frames by solving its equations of motion, 
which are given by 
\begin{align}
\nonumber
&\text{GT1} &&\text{GT2}\\
\frac{d\vec u}{d\zeta} &= \frac{\vec F}{v_z-c} &\frac{d\vec u}{d\tilde\zeta} &= \frac{\vec F}{v_z-c}\\
\frac{d s}{d\zeta} &= \frac{c}{v_z-c}      
&\frac{d \tilde s}{d\tilde\zeta} &= \frac{v_z}{v_z-c},
\end{align}
for spatial and time-like Galilean transformations, respectively. 
These effects of both transformations 
can be illustrated by solving a particle motion under the action of
a Gaussian pulse
\begin{align}
A_x &= a_0\cos(\phi)\exp\left(-\frac{\phi^2}{2\sigma^2}\right),
\end{align} 
where the phase $\phi$ is direct $\phi = (z-ct) = \zeta = \tilde\zeta$, 
and $a_0 =eA/mc= 1$  is the dimensionless amplitude. 
The action of spatial-like  GT1 and time-like GT2 transformations on a laser and a particle is illustrated in Figure~\ref{figGal}. The left panel shows the baseline motion in the laboratory frame, where a particle initially at rest is accelerated by a laser pulse propagating from left to right. After crossing the laser pulse, it remains at rest.  

The middle panel presents the situation in the Galilean frame, where the laser pulse remains stationary, and the particle is transported along the fast coordinate until it encounters the laser field. 
\replaced[id=R1]{After the interaction, the particle again transported along the fast coordinate.}{After the interaction, the particle remains at a constant $s$-coordinate. 
}

In the right panel, a time-like Galilean transformation results in a stationary laser pulse, while the particle passes through the pulse and, after the interaction, continues along a straight trajectory with constant velocity.

\section{Basic equations}
\label{BasicEqs}
We take the full set of Maxwell equations

\begin{equation}
\nabla\times{\bf B}=\frac{1}{c}\partial_{t}{\bf E}+\frac{4\pi}{c}{\bf j}
\end{equation}
and

\begin{equation}
\nabla\times{\bf E}=-\frac{1}{c}\partial_{t}{\bf B}
\end{equation}
together with the kinetic equation for the distribution function $f_{\alpha}$
for each particle class $\alpha$:

\begin{equation}
\partial_{t}f_{\alpha}\left({\bf r},{\bf p},t\right)+{\bf v}\cdot\partial_{{\bf r}}f_{\alpha}\left({\bf r},{\bf p},t\right)+q_{\alpha}\left[{\bf E}+\frac{1}{c}{\bf v\times B}\right]\cdot \partial_{\bf p} f_{\alpha}\left({\bf r},{\bf p},t\right)=0
\end{equation}
so that the self-consistent current density is

\begin{equation}
{\bf j}=\sum_{\alpha}q_{\alpha}\int d^{3}p{\bf v}f_{\alpha}\left({\bf r},{\bf p},t\right).
\end{equation}
Now, we apply the Galilean transformation \eqref{eq:Galilean1} to
this Vlasov plasma description. For simplicity, we exploit the standard
relativistic normalization. First, we select some basic frequency
$\omega_{0}$. This can represent the mean laser frequency, the
background plasma frequency, or any other physically relevant process.
Then, all fields are normalized like

\begin{equation}
\frac{e{\bf E}}{mc\omega_{0}}\rightarrow{\bf E},\,\,\,\,\frac{e{\bf B}}{mc\omega_{0}}\rightarrow{\bf B}.
\end{equation}
The lengths and time are normalized like

\begin{equation}
\frac{\omega_{0}}{c}{\bf r}\rightarrow{\bf r},\,\,\,\,\omega_{0}t\rightarrow t.
\end{equation}
The vacuum speed of light in these dimensionless variables becomes
$c=1$ and the Maxwell equations take the form of

\begin{equation}
\nabla\times{\bf B}=\partial_{t}{\bf E}+{\bf j},\label{eq:Afven}
\end{equation}

\begin{equation}
\nabla\times{\bf E}=-\partial_{t}{\bf B}.\label{eq:Faraday}
\end{equation}

\subsection{Maxwell equations}
First, we make a two-dimensional (2d) Fourier transformation of the Maxwell equations in the $(X,Y)$ plane:
\begin{eqnarray}
{\bf i} k_y E_{z}-\partial_z E_{y} & = & - \partial_{t}B_{x}, \nonumber \\
\partial_z E_{x}-{\bf i} k_x E_{z} & = & - \partial_{t}B_{y}, \nonumber \\
 {\bf i} k_x E_{y}  - {\bf i} k_y E_{x} & = & - \partial_{t}B_{z}, \label{eq:2dBz} \\
{\bf i} k_y B_{z}-\partial_z B_{y} & = &  \partial_{t}E_{x} + j_x, \nonumber \\
\partial_z B_{x}-{\bf i} k_x B_{z} & = &  \partial_{t}E_{y} + j_y, \nonumber \\
 {\bf i} k_x B_{y}  - {\bf i} k_y B_{x} & = &  \partial_{t}E_{z} + j_z. \label{eq:2dEz}
\end{eqnarray}
Taking linear combinations of transverse fields, we obtain values transported in the positive $Z-$direction (to the right) 
\begin{eqnarray}
R_x & = & E_x+B_y \label{eq:Rx}\\
R_y & = & E_y-B_x \label{eq:Ry}
\end{eqnarray}
and values transported in the negative $Z-$direction (to the left)
\begin{eqnarray}
L_x & = & E_x-B_y \label{eq:Lx}\\
L_y & = & E_y+B_x \label{eq:Ly}.
\end{eqnarray}
The characteristic equations for these combinations of transverse fields are as follows:
\begin{eqnarray}
\left( \partial_{t} + \partial_{z} \right)R_x & = & {\bf i} k_x E_{z} + {\bf i} k_y B_{z} - j_x, \nonumber \\
\left( \partial_{t} + \partial_{z} \right)R_y & = & {\bf i} k_y E_{z} - {\bf i} k_x B_{z} - j_y, \nonumber \\
\left( \partial_{t} - \partial_{z} \right)L_x & = & -{\bf i} k_x E_{z} + {\bf i} k_y B_{z} - j_x, \nonumber \\
\left( \partial_{t} - \partial_{z} \right)L_y & = & -{\bf i} k_y E_{z} - {\bf i} k_x B_{z} - j_y, \label{eq:2dCharacteristics}
\end{eqnarray}
Together with the equations on the longitudinal fields (\ref{eq:2dBz}) and (\ref{eq:2dEz}), these are the basic equations.

First, we apply the "spatial" transformation (\ref{eq:Galilean1}) and obtain the system:
\begin{eqnarray}
 \partial_{s}R_x & = & {\bf i} k_x E_{z} + {\bf i} k_y B_{z} - j_x, \label{eq:RxG1} \\
 \partial_{s}R_y & = & {\bf i} k_y E_{z} - {\bf i} k_x B_{z} - j_y, \label{eq:RyG1} \\
\left( \partial_{s} - 2 \partial_{\zeta} \right)L_x & = & -{\bf i} k_x E_{z} + {\bf i} k_y B_{z} - j_x, \label{eq:LxG1}\\
\left( \partial_{s} - 2 \partial_{\zeta} \right)L_y & = & -{\bf i} k_y E_{z} - {\bf i} k_x B_{z} - j_y,  \label{eq:LyG1}\\
  \left( \partial_{s} - \partial_{\zeta} \right)B_{z} & = & -{\bf i} k_x E_{y}  + {\bf i} k_y E_{x}  , \label{eq:BzG1} \\
\left( \partial_{s} - \partial_{\zeta} \right)E_{z}    & = &  {\bf i} k_x B_{y}  - {\bf i} k_y B_{x}  - j_z. \label{eq:EzG1}
\end{eqnarray}

For the "temporal" transformation (\ref{eq:Galilean2})  we obtain a slightly different system
\begin{eqnarray}
 \partial_{\tilde{s}}R_x & = & {\bf i} k_x E_{z} + {\bf i} k_y B_{z} - j_x, \label{eq:RxG2} \\
 \partial_{\tilde{s}}R_y & = & {\bf i} k_y E_{z} - {\bf i} k_x B_{z} - j_y, \label{eq:RyG2} \\
\left( -\partial_{\tilde{s}} - 2 \partial_{\tilde{\zeta}} \right)L_x & = & -{\bf i} k_x E_{z} + {\bf i} k_y B_{z} - j_x, \label{eq:LxG2}\\
\left( -\partial_{\tilde{s}} - 2 \partial_{\tilde{\zeta}} \right)L_y & = & -{\bf i} k_y E_{z} - {\bf i} k_x B_{z} - j_y,  \label{eq:LyG2}\\
- \partial_{\tilde{\zeta}}B_{z} & = & -{\bf i} k_x E_{y}  + {\bf i} k_y E_{x}  , \label{eq:BzG2} \\
- \partial_{\tilde{\zeta}}E_{z}    & = &  {\bf i} k_x B_{y}  - {\bf i} k_y B_{x}  - j_z. \label{eq:EzG2}
\end{eqnarray}

The main difference between these two transformations is the sign of $\partial_s-$derivative in the transport equation for $L_x,L_y$  as well as its presence in equations on the longitudinal fields (\ref{eq:BzG1}) and (\ref{eq:EzG1}). Keeping these derivatives over the slow coordinate $s$ prevents us from gaining any advantage over the untransformed Maxwell equations. However, we are interested in slowly evolving drivers (either particle bunches or lasers) and waves running at nearly the speed of light in the forward direction. For such waves, $\partial_s \ll \partial_\zeta$. Thus, we can neglect the small terms $\partial_s$ in all equations, where it appears together with the fast coordinate derivative $\partial_\zeta$. The transformed system of the Maxwell equations takes on the universal form
\begin{eqnarray}
 \partial_{s}R_x & = & {\bf i} k_x E_{z} + {\bf i} k_y B_{z} - j_x, \label{eq:RxGu} \\
 \partial_{s}R_y & = & {\bf i} k_y E_{z} - {\bf i} k_x B_{z} - j_y, \label{eq:RyGu} \\
  2 \partial_{\zeta} L_x & = &{\bf i} k_x E_{z} - {\bf i} k_y B_{z} + j_x, \label{eq:LxGu}\\
 2 \partial_{\zeta} L_y & = & {\bf i} k_y E_{z} + {\bf i} k_x B_{z} + j_y,  \label{eq:LyGu}\\
 \partial_{\zeta}B_{z} & = & {\bf i} k_x E_{y}  - {\bf i} k_y E_{x}  , \label{eq:BzGu} \\
 \partial_{\zeta} E_{z}    & = &  -{\bf i} k_x B_{y}  + {\bf i} k_y B_{x}  + j_z, \label{eq:EzGu}
\end{eqnarray}
valid for both Galilean transformations.

\subsection{Validity of approximation}

Let us consider the influence of backward propagating terms on a linear solution of the Maxwell equations. 
In 1D, a linearly polarized electromagnetic wave propagating along the $z$ coordinate can be defined by the field components $\vec E = \{E_x, 0,0\}$ and $\vec B = \{0, B_y, 0\}$. 
\added[id=R2]{With this choice of polarization,} the full Maxwell system can be effectively reduced to a pair of equations for left- and right-propagating characteristics. 

\begin{align}
\{L_x, R_x\} = \{E_x-B_y,\; E_x+B_y\} = \{L,R\}.
\end{align}

In vacuum these characteristics are decoupled. \added[id=R2]{To incorporate plasma effects, we adopt a simplified model in which the interaction is described through the charge current density} \cite{Spranlge:2002}
\begin{align}
\frac{\partial \vec J}{\partial t} +\nu_e \vec J = \omega^2_p\left( 1+ \frac{\delta n_e}{n_e}\right)\vec E, 
\end{align}
where $\delta n_e$ is the density perturbation due to the longitudinal wakefield and $\nu_e$ is the electron collision frequency. 
\added[id=R2]{For the purpose of the present analysis, we neglect collisions and density perturbations, i.e.} 
$\nu_e = 0$, $\delta n_e \simeq 0$. 
\added[id=R2]{Under these assumptions, the equation for the current density can be written as}

\begin{align}
\text{GT1:}&\\
\partial_sR &= -J\\
(\partial_s-2\partial_\zeta)L &= -J\\
\quad  \frac{\partial J}{\partial s} - c\frac{\partial J}{\partial \zeta} &= 
\omega^2_p E
\end{align}

\begin{align}
\text{GT2:}&\\
\partial_sR &= -J\\
(-\partial_s-2\partial_\zeta)L &= -J\\
\quad\quad  -c\frac{\partial J}{\partial \tilde \zeta} &= \omega^2_p E\\
\end{align}

\added[id=R2]{The analytical dispersion relations can be obtained using the Fourier ansatz}
\begin{align}
\begin{bmatrix}
L\\
R\\
J
\end{bmatrix} = 
\begin{bmatrix}
\hat L\\
\hat R\\
\hat J
\end{bmatrix}\exp(i\kappa\cdot\zeta - i\omega\cdot s).
\end{align}

\added[id=R2]{After applying the Fourier transform, the systems for the two cases can be written in the matrix form}

\begin{align}
\text{GT1:}
\begin{bmatrix}
-i\omega &-i\omega &1\\
\cancel{-i\omega}-2i\kappa & \cancel{i\omega}+2i\kappa & 1\\
-\omega^2_p &0&\cancel{-i\omega}-i\kappa
\end{bmatrix}
\begin{bmatrix}
\hat E\\
\hat B\\
\hat J
\end{bmatrix}=0
\label{matGT1}
\end{align}

\begin{align}
\text{GT2:}
\begin{bmatrix}
-i\omega &-i\omega &1\\
\cancel{i\omega}-2i\kappa & -(\cancel{i\omega}-2i\kappa) & 1\\
-\omega^2_p &0&-i\kappa
\end{bmatrix}
\begin{bmatrix}
\hat E\\
\hat B\\
\hat J
\end{bmatrix}=0
\label{matGT2}
\end{align}

\added[id=R1]{These linearized systems correspond to the respective transformations and retain all backward–propagating contributions. In the universal reduced system  Equations~{(\ref{eq:RxGu}-\ref{eq:EzGu})} these terms do not appear; therefore they are indicated by crossed-out entries in the matrices. In the following, we show that the full dispersion relations, when evaluated in the asymptotic limit $\partial_s L \to 0$, reduce to the same dispersion relation as that obtained in the simplified system where these terms are explicitly discarded.}

The dispersion relations are obtained by imposing $\mbox{det}=0$. 
This procedure yields the following expressions

\begin{align}
\text{GT1:}\qquad 
F(\omega,\kappa) &= -2\kappa^2\omega-\kappa\omega^2+\kappa\omega^2_p+\omega\omega^2_p
\\
\text{GT2:}\qquad F(\omega,\kappa) &= \kappa(-2\kappa\omega+\omega^2+\omega^2_p)
\end{align}

The explicit solutions of these systems given below

\begin{align}
\text{GT1:}\quad \kappa_{1,2} &= \frac{-\omega^2+\omega^2_p\pm\sqrt{\omega^4-6\omega^2\omega^2_p+\omega^4_p}}{4\omega}
\label{Det}
\\
\text{GT2:}\quad \kappa_{1,2} &= \left\{0,\frac{\omega^2+\omega^2_p}{2\omega}\right\}
\end{align}

\added[id=R2]{Both dispersions can be reduced to a 
universal form by taking the asymptotic limit} 
$\partial_s L\rightarrow 0$ which means crossed terms in $\cancel{i\omega} \rightarrow 0$.  
\begin{align}
\kappa^a_{1,2} = \frac{\omega^2_p\pm \omega_p\sqrt{4\omega^2+\omega^2_p}}{4\omega}
\end{align}

\begin{figure}[ht]
\includegraphics[width=14cm]{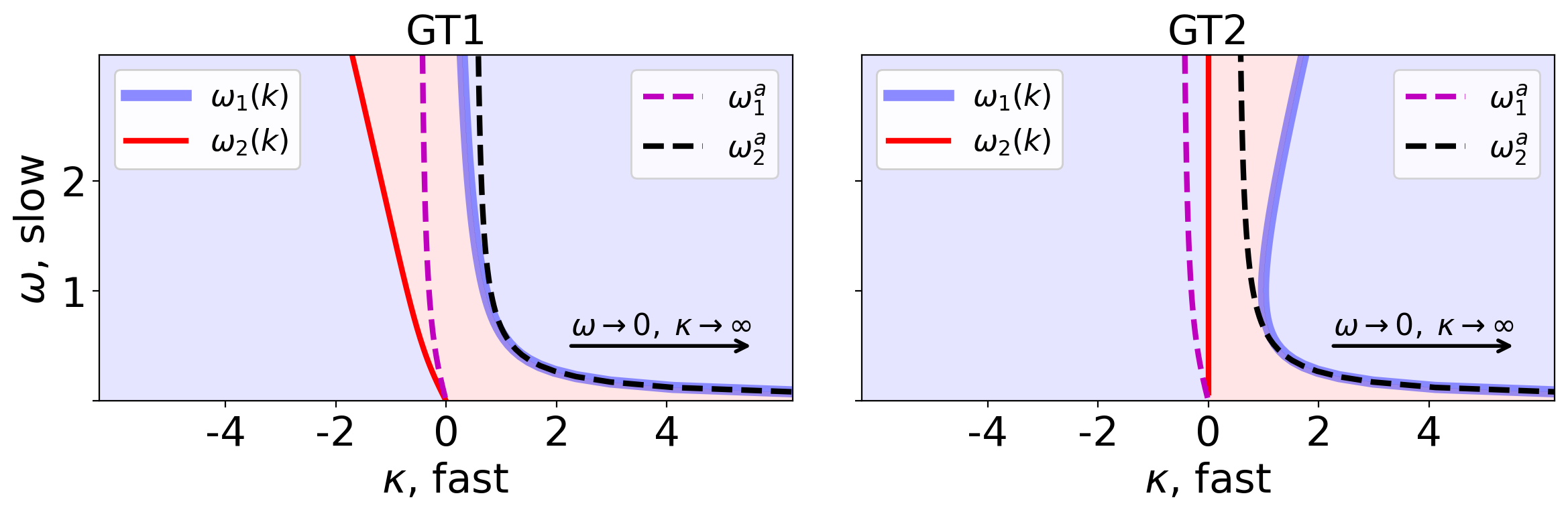}
\caption{
Dispersion relations corresponding to the GT1 and GT2 formulations compared with the universal dispersion (\ref{fu}). The curves are shown for $\omega_p^2=1$. The close agreement confirms that neglecting the backward propagating mode is justified for sufficiently large wave numbers. 
}
\label{Disp}
\end{figure}

This coincides with the solution 
of universal dispersion relation obtained by 
neglecting $\partial_sL=0$ as stated by Equations~{(\ref{eq:RxGu}-\ref{eq:EzGu})}
\begin{align}
F_u(\omega,\kappa) = -4\kappa^2\omega+2\kappa\omega^2_p+\omega\omega^2_p = 0.
\label{fu}
\end{align}
Therefore asymptotic limit $\kappa^a(\omega) = \kappa^u(\omega)$. The full and asymptotic dispersions are plotted  in Fig.~\ref{Disp}, where full dispersion corresponds to $F(\omega,\kappa) = 0$.

The dispersion diagrams reveal two distinct branches corresponding to forward- and backward-propagating modes. For sufficiently large $\kappa$, the branches are well separated, and the forward-propagating mode of primary interest (depicted as a thick blue line in both plots) closely follows the asymptotic solution (black dashed line), which approaches the universal dispersion relation, indicating that the reduced model accurately captures the wave dynamics in this regime.

\section{Finite-difference spectral solver in 3d}
\subsection{Finite-difference spectral scheme in 3d}
We denote the large step along the slow coordinate $s$ by $\Delta$,
and the fine step along the fast coordinate $\zeta$ by $h$. 
Numerical indexing follows the convention in which the upper index refers to steps
along $s$, and the lower index corresponds to steps along $s$.
We assume that the solution is known at $s=s^{0}$, and that at $s=s^{1}=s^{0}+\Delta$,
all quantities are known to the right of $\zeta=\zeta_{0}$ (i.e., for
$\zeta>\zeta_{0})$. With this setup, the numerical scheme can be formulated
as follows:

\begin{eqnarray}
\frac{R_{x0}^{1}-R_{x0}^{0}}{\Delta}&= & \frac{1}{2}\left(ik_{x}E_{z0}^{1}+ik_{x}E_{z0}^{0}+ik_{y}B_{z0}^{1}+ik_{y}B_{z0}^{0}\right)
\nonumber \\
& & -\frac{j_{x-1/2}^{1}+j_{x-1/2}^{0}+j_{x+1/2}^{1}+j_{x+1/2}^{0}}{4} \label{RxN}  \\
\frac{R_{y0}^{1}-R_{y0}^{0}}{\Delta} & =&\frac{1}{2}\left(ik_{y}E_{z0}^{1}+ik_{y}E_{z0}^{0}-ik_{x}B_{z0}^{1}-ik_{x}B_{z0}^{0}\right) \nonumber  \\
& &-
\frac{1}{4} \left(
j_{y-1/2}^{1}+j_{y-1/2}^{0}+
j_{y+1/2}^{1}+j_{y+1/2}^{0}\right)\label{RyN}  \\
2\frac{L_{x1}^{1}-L_{x0}^{1}}{h}&= & ik_{x}\frac{E_{z1}^{1}+E_{z0}^{1}}{2}-ik_{y}\frac{B_{z0}^{1}+B_{z1}^{1}}{2}+j_{x+1/2}^{1}  \\
2\frac{L_{y1}^{1}-L_{y0}^{1}}{h}&= & ik_{y}\frac{E_{z1}^{1}+E_{z0}^{1}}{2}+ik_{x}\frac{B_{z0}^{1}+B_{z0}^{0}}{2}+j_{y+1/2}^{1}  \label{LyN} \\
\frac{B_{z1}^{1}-B_{z0}^{1}}{h}&=&ik_{x}\frac{E_{y0}^{1}+E_{y1}^{1}}{2}-ik_{y}\frac{E_{x0}^{1}+E_{x1}^{1}}{2}   \\
\frac{E_{z1}^{1}-E_{z0}^{1}}{h}&=&-ik_{x}\frac{B_{y0}^{1}+B_{y1}^{1}}{2}-ik_{y}\frac{B_{x0}^{1}+B_{x1}^{1}}{2} +j_{z+1/2}^{1} \label{eqall}
\end{eqnarray}
This scheme is implicit, but can be easily solved for unknown
fields  ${\bf E}_{0}^{1},{\bf B}_{0}^{1}$  at the position $\zeta=\zeta^0$, $s=s^1$.  
\added[id=R2]{To obtain the fields at the next level $s$, we start marching along $\zeta$ from the right boundary of the simulation domain. } 

Currents $j_{x-1/2}^{1}$ and $j_{y-1/2}^{1}$ have to be found using a predictor-corrector procedure.
However, predictions of these currents affect only the updates along
the slow coordinate $s$ in Eqs. \eqref{RxN} and \eqref{RyN}.
At the next leap-frog step along $\zeta$, these currents are accurately recalculated.

The analytical solution to system (\ref{RxN})-(\ref{eqall}) is given in \ref{DispAppSolution}.

\subsection{Numerical dispersion}
\label{disp1d}
Now we derive the numerical dispersion relation from a 
discretized version of the universal Maxwell system written as
\begin{align}
\frac{R^k_{n} - R^{k-1}_n}{\Delta } &= 
-\frac{1}{4}\left(J^k_{n-1/2} + J^{k-1}_{n-1/2}
+J^k_{n+1/2} + J^{k-1}_{n+1/2}\right)\\
-2\frac{L^k_{n+1}-L^k_n}{h} &= -J^k_{n+1/2}\\
-\frac{J^k_{n+\frac{1}{2}}-J^k_{n-\frac{1}{2}}}{h} &= 
\omega^2_p E^k_n.
\label{num1d}
\end{align}

\added[id=R2]{While the time discretization of the $R$ equation corresponds to a trapezoidal Crank–Nicolson–type integration, the overall scheme is not a classical Crank–Nicolson method. Instead, it represents a semi-implicit characteristic box scheme formulated in the $(s,\zeta)$ coordinate system.}

\begin{figure}[ht]
\includegraphics[width=12cm]{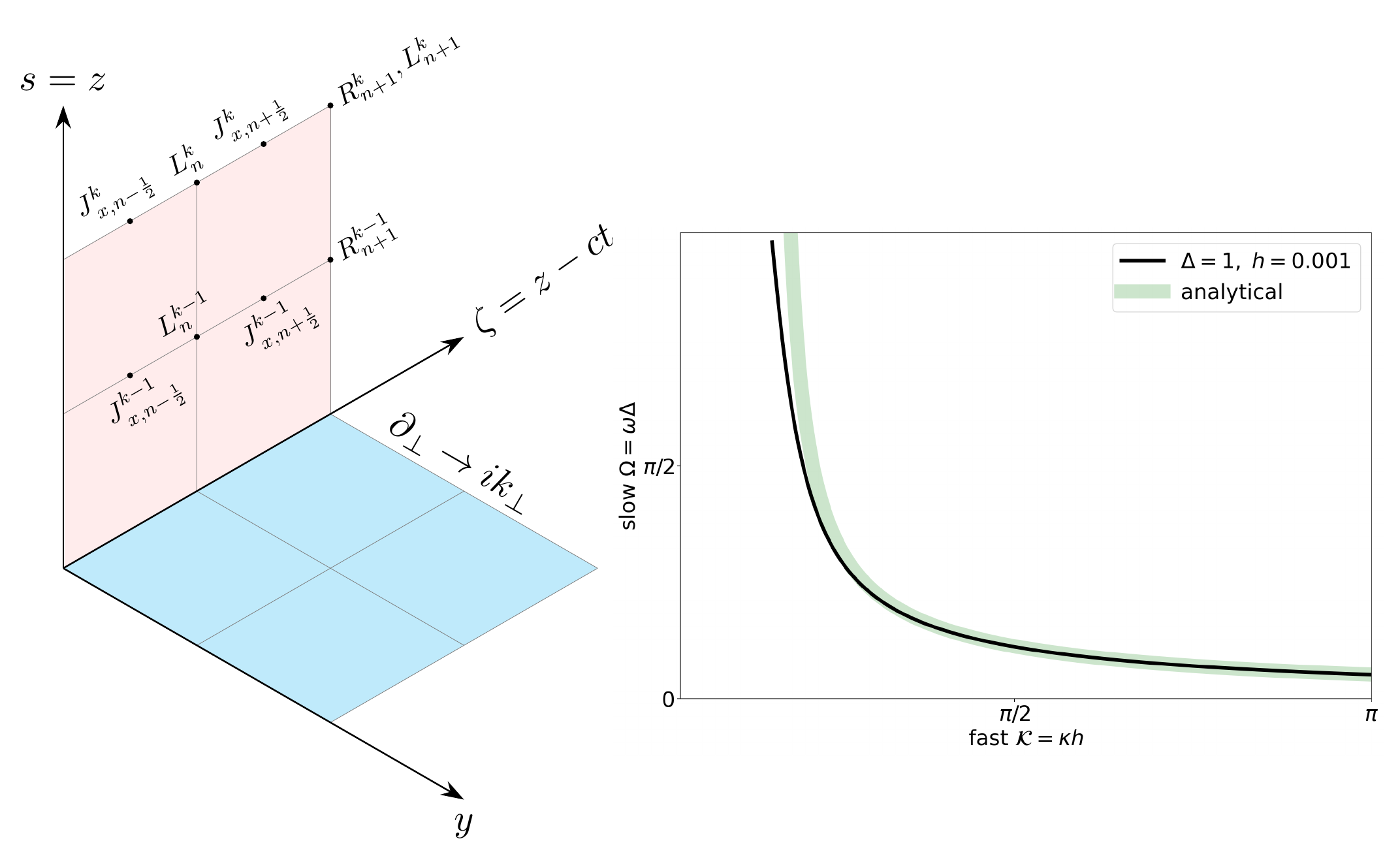}
\caption{
Left: grid stencil of the semi-implicit scheme used for the discretization of the universal system. 
Right: comparison of the numerical dispersion relation with the analytical dispersion (\ref{fu}). The close agreement confirms that the discretization introduces only weak numerical dispersion.
}
\label{tau}
\end{figure}

The corresponding grid stencil is depicted in Figure~\ref{tau}. The current density is defined at half steps according to the leap-frog staggering commonly used in particle-in-cell methods.

The dispersion introduced by finite differences can be calculated using the discrete Fourier ansatz
\begin{align}
R^k_n = \hat R\exp(-i\omega s^k + i\kappa\zeta_n).
\label{dft}
\end{align}

\added[id=R2]{Substituting the discrete Fourier ansatz (\ref{dft}) into the discretized system (\ref{num1d}) and eliminating the field amplitudes yields the following numerical dispersion relation}

\begin{align}
\tan\left(\frac{\omega\Delta }{2}\right) = -\frac{\omega^2_p\Delta h\sin(\kappa h)}
{\omega^2_p h^2+ 8\cos(\kappa h) -8}
\end{align}

\added[id=R2]{The scheme is second-order accurate. Indeed, using the Taylor expansions of the trigonometric functions}

\begin{align}
\tan(x) = x+\mathcal O(x^3),\quad 
\sin(x) = x + \mathcal O(x^3),
\end{align}

\added[id=R2]{one recovers the analytical dispersion relation with second-order accuracy} $\propto \mathcal O(\Delta^2,h^2)$.
In the long-wavelength limit $\kappa h\ll1$ and $\omega\Delta \ll 1$, the numerical dispersion reduces to the analytical form obtained from (\ref{fu})
\begin{align}
\omega = \frac{2\kappa\omega^2_p}{4\kappa^2 - \omega^2_p},
\end{align}
up to corrections of order $\mathcal O((\kappa\Delta\zeta)^2,(\omega\Delta)^2)$.
It is also worth noting that the resulting dispersion relation does not impose a classical CFL-type restriction on the step $\Delta$, which reflects the semi-implicit character of the scheme. A practical stability estimate can be obtained from the most
unstable mode $\sin(\kappa h/2)=1$:
\[
\left|
\frac{\omega_p^2\Delta h}{4}
\sin\left(\frac{\kappa h}{2}\right)
\right|\le1,
\]
which leads to the CFL-like stability condition,
\begin{align}
\Delta\cdot h \le \frac{4}{\omega_p^2}.
\end{align}
This dispersion is plotted in Figure~\ref{DispApp} and compared 
with analytical solution for the system nomalized by plasma time scale $\omega_0=\omega_p$, see Section~(\ref{BasicEqs}).
\added[id=R2]{The close agreement between the numerical and analytical dispersion relations also indicates that the scheme does not introduce artificial slowing of the phase velocity, which helps suppress numerical Cherenkov–type artifacts often encountered in standard FDTD discretizations.}
A more complete derivation of the dispersion properties can be found in~\ref{DispApp}.

\section{Transformed particle-in-cell method and particle trapping}
\label{sec1}

Particle-in-Cell codes push the so-called ``macro-particles" which represent "clumps" of physical particles, all moving along the same trajectory. This is mathematically possible, because we
solve the Vlasov equation along characteristics. The Vlasov equation
is a typical transport equation on an "incompressible fluid" in the 6-dimensional phase space. As a consequence, the distribution function $f(t,{\bf r},{\bf p})$ is preserved along its characteristics. 

The Vlasov equation is

\begin{equation}
\partial_{t}f(t,{\bf r},{\bf p})+{\bf v}\cdot\partial_{{\bf r}}f(t,{\bf r},{\bf p})+{\bf F}\cdot\partial_{{\bf p}}f(t,{\bf r},{\bf p})=0,
\end{equation}
where $\bf F$ is the force that acts on the particles. 

First, we transform it using the "spatial" Galilean 
coordinates (\ref{eq:Galilean1}). Changing to the $(s,\zeta)$-coordinates, it becomes

\begin{eqnarray}
& & \partial_{s}f(s,\zeta,{\bf r_{\perp}},{\bf p})   +    \left(v_{z}-1\right)\partial_{\zeta}f(s,\zeta,{\bf r_{\perp}},{\bf p})  \nonumber \\
 & & + {\bf v_{\perp}\cdot}\partial_{{\bf r_{\perp}}}f(s,\zeta,{\bf r_{\perp}},{\bf p}) + {\bf F}\cdot\partial_{{\bf p}}f(s,\zeta,{\bf r_{\perp}},{\bf p})=0  \label{eq:Vlasov1}
\end{eqnarray}
Let us parameterize the characteristics with an additional parameter
$\eta$. Then the equations on the characteristics are as follows:

\begin{equation}
\frac{ds}{d\eta}=1,\,\,\,\frac{d\zeta}{d\eta}=v_{z}-1,\,\,\,\frac{d{\bf r_{\perp}}}{d\eta}={\bf v_{\perp}},\,\,\,\frac{d{\bf p}}{d\eta}={\bf F} 
\end{equation}
Because $v_{z}<1$, the dependence of the parameter $\eta$ on the
fast variable $\zeta$  is single-valued, and we can choose $\zeta$ as
the new parameter. We then have the following.

\begin{equation}
\frac{ds}{d\zeta}=\frac{1}{v_{z}-1},\,\,\,\frac{d{\bf r_{\perp}}}{d\zeta}=\frac{{\bf v_{\perp}}}{v_{z}-1},\,\,\,\frac{d{\bf p}}{d\zeta}=\frac{{\bf F}}{v_{z}-1}.
\end{equation}
The grid step $\Delta$ along the coordinate $s$ corresponds to the
step along time: $c \delta t=\Delta$. The grid step $h$ along the fast changing coordinate $\zeta$ is small: $h\ll\Delta$.

We push particles along the coordinate $\zeta$ during the time step $\Delta$. This means that each particle has its initial coordinate $s^{0}$ and we continue the push in $\zeta$ according to 

\begin{equation}
s_{i+1}=s_{i}+\frac{1}{v_{z}-1}d\zeta \label{eq:trap1}
\end{equation}
until the particle reaches $s^{1}=s^{0}+\Delta$, or it leaves the simulation domain at the left boundary in $\zeta$. The step $-d\zeta$ is then the smallest one between $h$ and $\Delta\left(1-v_{z}\right)$.

The "spatial" Galilean transformation is the most general one, and it can describe the trapping as closest to a full PIC code as possible. However, the trapping condition (\ref{eq:trap1}) means that we have to stop pushing even physically untrapped particles when the step $\Delta$ along the slow coordinate $s$ is smaller than the simulation domain length $L$. This leads to artificial "stitches" in the particle density and eventual artifacts in the fields.

Another approach is the "temporal" Galilean transformation (\ref{eq:Galilean2}). This was not possible in the previous quasi-static codes. The Vlasov equation transformed to the $(\tilde{s},\tilde{\zeta})$-coordinates, it becomes
\begin{eqnarray}
& & v_z \partial_{\tilde{s}}f(\tilde{s},\tilde{\zeta},{\bf r_{\perp}},{\bf p})  +  \left(v_{z}-1\right)\partial_{\tilde{\zeta}}f(\tilde{s},\tilde{\zeta},{\bf r_{\perp}},{\bf p})  \nonumber \\
 &  &  + {\bf v_{\perp}\cdot}\partial_{{\bf r_{\perp}}}f(\tilde{s},\tilde{\zeta},{\bf r_{\perp}},{\bf p}) + {\bf F}\cdot\partial_{{\bf p}}f(\tilde{s},\tilde{\zeta},{\bf r_{\perp}},{\bf p})=0 \label{eq:Vlasov2}
\end{eqnarray}
The only difference between (\ref{eq:Vlasov1}) and (\ref{eq:Vlasov2})  is the velocity $v_z$ in front of the derivative along the slow coordinate $s$ in equation (\ref{eq:Vlasov2}) .  Let us parameterize the characteristics with an additional parameter $\tilde{\eta}$. Then the equations on the characteristics are as follows:

\begin{equation}
\frac{d\tilde{s}}{d\tilde{\eta}}=v_z,\,\,\,\frac{d\tilde{\zeta}}{d\tilde{\eta}}=v_{z}-1,\,\,\,\frac{d{\bf r_{\perp}}}{d\tilde{\eta}}={\bf v_{\perp}},\,\,\,\frac{d{\bf p}}{d\tilde{\eta}}={\bf F}
\end{equation}
Again, because $v_{z}<1$, the dependence of the parameter $\tilde{\eta}$ on the
fast variable $\tilde{\zeta}$  is single-valued, and we can choose $\zeta$ as
the new parameter. We then have the following.

\begin{equation}
\frac{d\tilde{s}}{d\tilde{\zeta}}=\frac{v_z}{v_{z}-1},\,\,\,\frac{d{\bf r_{\perp}}}{d\tilde{\zeta}}=\frac{{\bf v_{\perp}}}{v_{z}-1},\,\,\,\frac{d{\bf p}}{d\tilde{\zeta}}=\frac{{\bf F}}{v_{z}-1}.
\end{equation}
The grid step $\Delta$ along the coordinate $s$ corresponds to the
step along the propagation distance. The grid step $h$ along the fast changing coordinate $\zeta$ is small: $h\ll\Delta$.

We push particles along the coordinate $\zeta$ during the time step $\Delta$. This means that each particle has its initial coordinate $s^{0}$ and we continue the push in $\zeta$ according to 

\begin{equation}
\tilde{s}_{i+1}=\tilde{s}_{i}+\frac{v_z}{v_{z}-1}d\tilde{\zeta} \label{eq:trap2}
\end{equation}
until the particle reaches $\tilde{s}^{1}=\tilde{s}^{0}+\Delta$, or it leaves the
simulation domain at the left boundary in $\zeta$. The step $-d\zeta$ is then the smallest one between $h$ and $\Delta\left(1-v_{z}\right)/v_z$. 

The particles, which have reached the next $s-$level during integration
along the fast $\zeta-$coordinate, remain inside the simulation box
and become trapped. \textit{This algorithm allows us to self-conistently simulate the 
self-trapping process for the first time} in codes
that use the Galilean transformation \eqref{eq:Galilean1}. This was
not possible in the standard quasi-static codes.


\subsection{Current Deposition}

We develop a full PIC code, where we do not distinguish between ``beam''
and ``jet'' numerical particles used in the quasi-static formulation.
All particles are treated uniformly. However, we are going to have
particles with various initial conditions.

First, we have ``fresh'' macroparticles that our simulation domain
``sweeps over" at every step in $s$. One such macroparticle carries
the total charge (physical ``weight'')

\begin{equation}
Q_{f}=qnh_{x}h_{y}\Delta,
\end{equation}
where $h_{y},h_{z}$ are the transverse grid steps. Here, $n$ is the density of the background particles.

Second, we may have particles, which we initialize within the simulation
domain. These may represent the ``beam'' or the ``driver''. One such macro-particle carries the total charge

\begin{equation}
Q_{b}=qnh_{x}h_{y}h,
\end{equation}
The macroparticle generates the current density ${\bf j}$ on the grid:

\begin{equation}
{\bf j}=\frac{Q}{h_{x}h_{y}h}{\bf v}\frac{ds}{\Delta},
\end{equation}
where 

\begin{equation}
ds=\frac{dt}{v_{z}-1}
\end{equation}
is the time the particle spends inside one cell.

Consider a ``fresh'' particle that is not ``trapped''.
In this case, $dt=-h$ and we have for the current deposition

\begin{equation}
{\bf j}=qn\frac{{\bf v}}{1-v_{z}}.
\end{equation}
For a ``trapped'' or ``beam'' particle with $\Delta\left(1-v_{z}\right)<h$,
one has to use $dt=-\Delta\left(1-v_{z}\right)$ and we obtain

\begin{equation}
{\bf j}=qn{\bf v}.
\end{equation}
Thus, we can smoothly process all the particles inside the simulation
domain. However, it is advantageous to integrate the trapped particles
along the slow $s$ coordinate, because numerically this is more
accurate. In fact, integrating them along the fast coordinate $\zeta$
would require knowing their $v_{z}$ before and after the push. This
makes the push scheme implicit and complicated. Integration along
$s$ avoids this complication.

\section{Example simulation}
In this section, we provide an example simulation for wake-field generation and particle trapping using the full three-dimensional (3D) GEM-PIC code. We compare these results with the quasi-3D cylindrical electromagnetic PIC code FBPIC \cite{FBPIC}. 

\begin{figure}
\includegraphics[width=7cm]{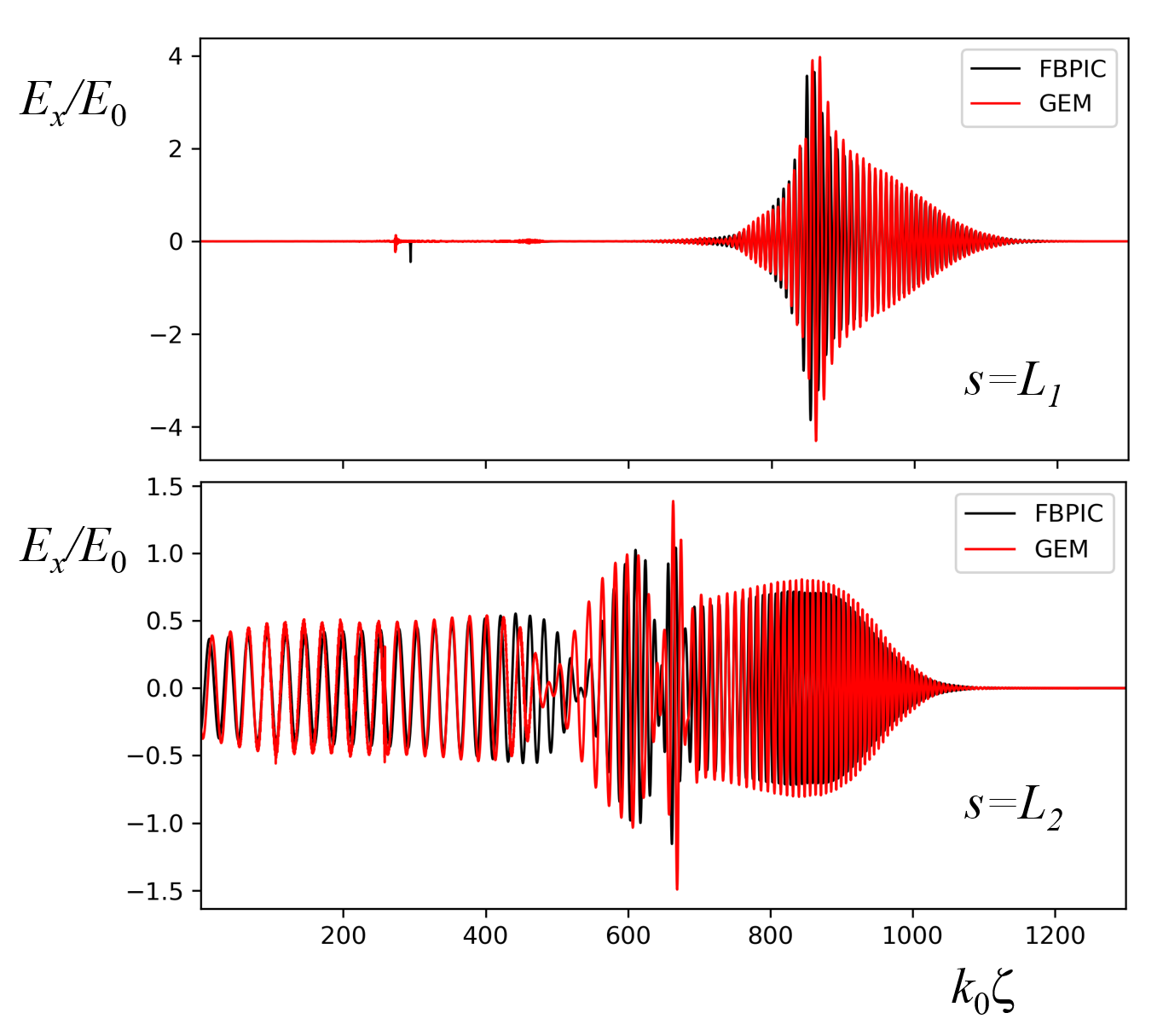}
\caption{On-axis laser field obtained in FBPIC and GEM-PIC simulations after the propagation distance of $L_1=15.9$~cm and $L_2=31.8$~cm.}
\label{Fig:ExOnAxis}
\end{figure}

We simulate a laser pulse of 25~J energy with a Gaussian profile

\begin{equation}
{\bf E}(t,{\bf r_\perp },z)={\bf e_x} a_0 \cos ( \phi) \exp (-\phi^2/T^2)\exp (-r_\perp ^2/R^2), \label{E_laser}
\end{equation} 
where $\phi=z-z_0 -ct$ is the laser phase,  $a_0=2$  is the normalized laser amplitude, $T=50.9$~fs is the  duration of the pulse and $R=22.9~\mu$m is the radius of the focal spot.  the laser wavelength $\lambda=800$~nm. 

\begin{figure}
\includegraphics[width=10cm]{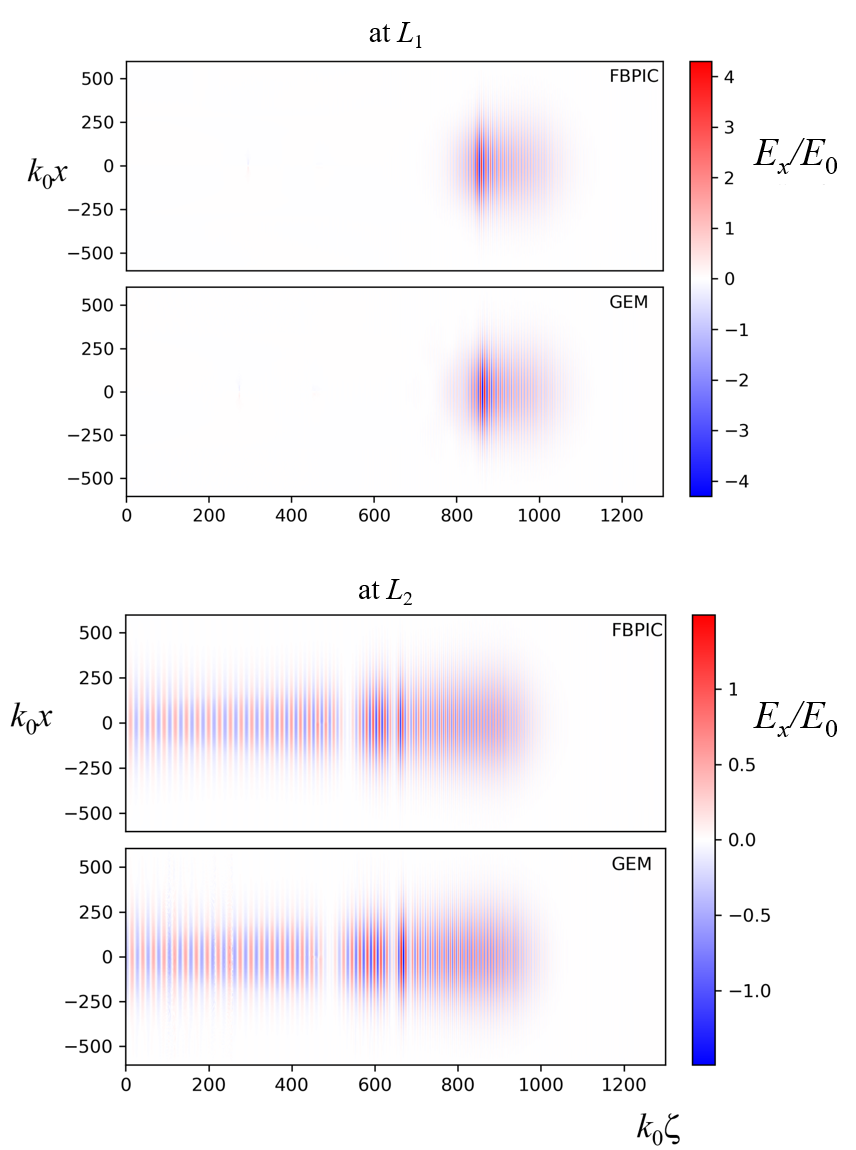}
\caption{2d distributions of the laser field $E_x$ obtained in FBPIC and GEM-PIC simulations.}
\label{Fig:Ex}
\end{figure}

The plasma was a preionized hydrogen  with a Gaussian channel. The radial profile of the density of electrons $n_e$ and protons $n_p$  is

\begin{equation}
n_e({\bf r_\perp })=n_p({\bf r_\perp })=n_0  \left[ 1+\frac{n_1-n_0}{n_0}\exp \left(-{\bf r_\perp } ^2/R_{\rm ch}^2\right) \right], \label{n_plasma}
\end{equation}
where  $n_0=5\times10^{-17}~1/{\rm cm^3}$ is the density of the background and $n_1=1.25\times10^{-17}~1/{\rm cm^3}$ is the density at the bottom of the channel. The channel radius is $R_{\rm ch}=76.4~\mu$m. The accelerating distance (the total plasma length) was $L_{\rm acc}=31.8$~cm.

To facilitate electron trapping, nitrogen doping is used \cite{Bella:2023}.  Nitrogen is assumed initially to be unionized with atomic density $n_{N}=0.02n_p$. Doping is placed at $0.63~\rm{cm} <z<1.27~\rm{cm}$ along the propagation distance.

The simulation window $X\times Y \times Z$ was $150 \times 150 \times 180~\mu{\rm m}^3$. The GEM-PIC simulation sampled this domain with the transverse grid steps $h_x=h_y=1~\mu{\rm m}$. The longitudinal grid step was flexible. In the backward third of the simulation domain, where the high amplitude wake field is located, the grid step was $h_z=6.3~{\rm nm}$. This high resolution was required to stably resolve the backward wall of the bubble located here. In the forward part of the simulation domain, where the laser pulse was propagating, the longitudinal grid step was $h_z=50~{\rm nm}$. The step along the propagation distance (the slow $s$ coordinate) varied from $\Delta=32~\mu{\rm m}$ at the beginning of the simulation to $\Delta=64~\mu{\rm m}$  after the laser has passed the nitrogen doping region. GEM-PIC used 4 numerical macro-atoms per cell for hydrogen and 4 macro-atoms per cell for nitrogen.

The FBPIC simulation used the cylindrical simulation window $r_\perp \times Z$ of $150 \times 180~\mu{\rm m}^2$ with 2 azimuthal modes. The radial step was $h_r=0.4~\mu{\rm m}$ and the longitudinal grid step was 64 cells per laser wavelength. The FBPIC simulation was performed in a Lorentz-boosted frame \cite{WarpX} with $\gamma=6$.

The GEM-PIC simulation took 12 hours on 432 cores of Intel(R) Xeon(R) CPU E5-2697 v2 @ 2.70GHz at a local cluster. The FBPIC simulation was run for 12 hours on 16 NVIDIA Tesla Ampere A100 GPUs at the JUPITER supercomputer, Jülich Supercomputing Centre.

We compare here results of both codes after the propagation distances $L_1=15.9~$cm (half propagation distance) and $L_2=31.8$~cm (end of simulation).  

 Fig.~\ref{Fig:ExOnAxis} compares the on-axis field $E_z$  (the laser polarization) at the positions $L_1$ and $L_2$ obtained in both codes. We observe a good agreement in the nonlinear evolution of the laser pulse envelope as well as the field phase.

 Fig.~\ref{Fig:Ex} shows the two-dimensional distributions of the laser field $E_x$ at the positions $L_1$ and $L_2$. Close to the end of simulation, $s=L_2$, the laser redshift was so strong that the long wavelength tail fills all the bubble.

\begin{figure}
\includegraphics[width=8cm]{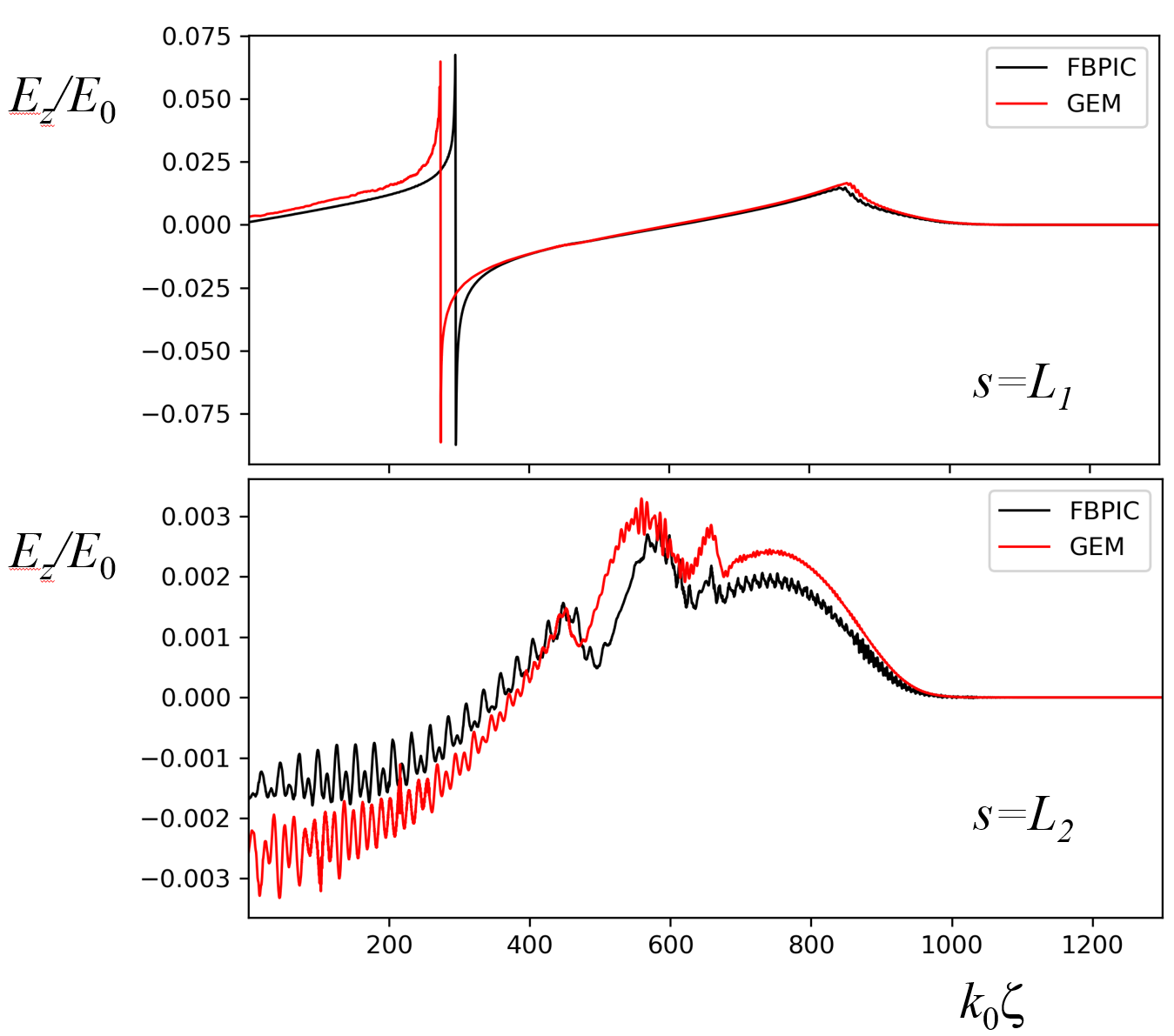}
\caption{On-axis wake field generated in plasma as obtained in FBPIC and GEM-PIC simulations after the propagation distance of $L_1=15.9$~cm and $L_2=31.8$~cm.}
\label{Fig:WakeOnAxis}
\end{figure}

\begin{figure}
\includegraphics[width=10cm]{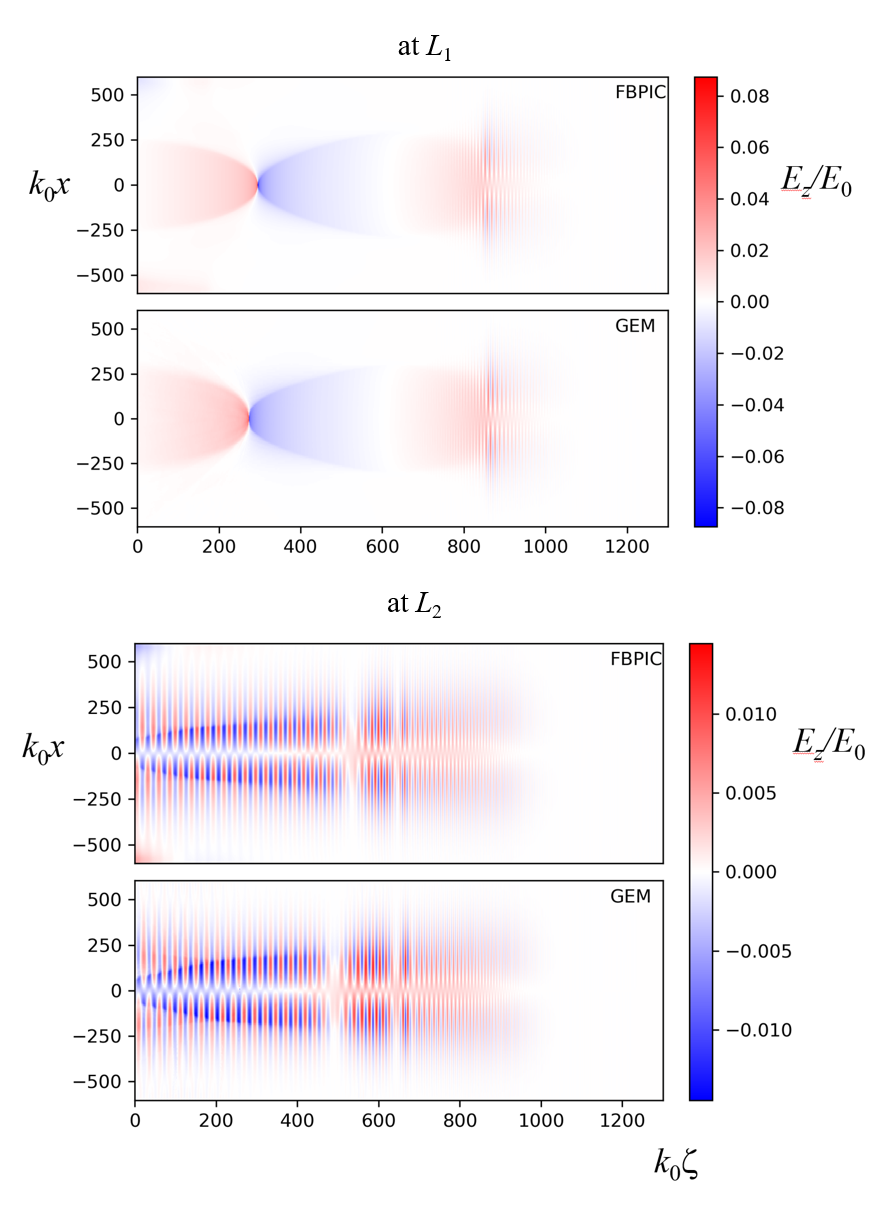}
\caption{2d distributions of the accelerating wakefield $E_z$ obtained in FBPIC and GEM-PIC simulations.}
\label{Fig:Wake2D}
\end{figure}

 Fig.~\ref{Fig:WakeOnAxis} shows the on-axis wakefield generated by the laser at the positions $L_1$ and $L_2$ obtained in both codes.  The plasma bubble is slightly longer in the GEM-PIC simulation because the laser pulse has a bit higher intensity at this position. Most probably, this difference is caused by a slightly more efficient absorption of the diffracted/scattered laser part at the lateral boundaries of the simulation box in FBPIC.  The two-dimensional distribution of the longitudinal accelerating field is shown in Fig.~\ref{Fig:Wake2D}.

\begin{figure}
\includegraphics[width=10cm]{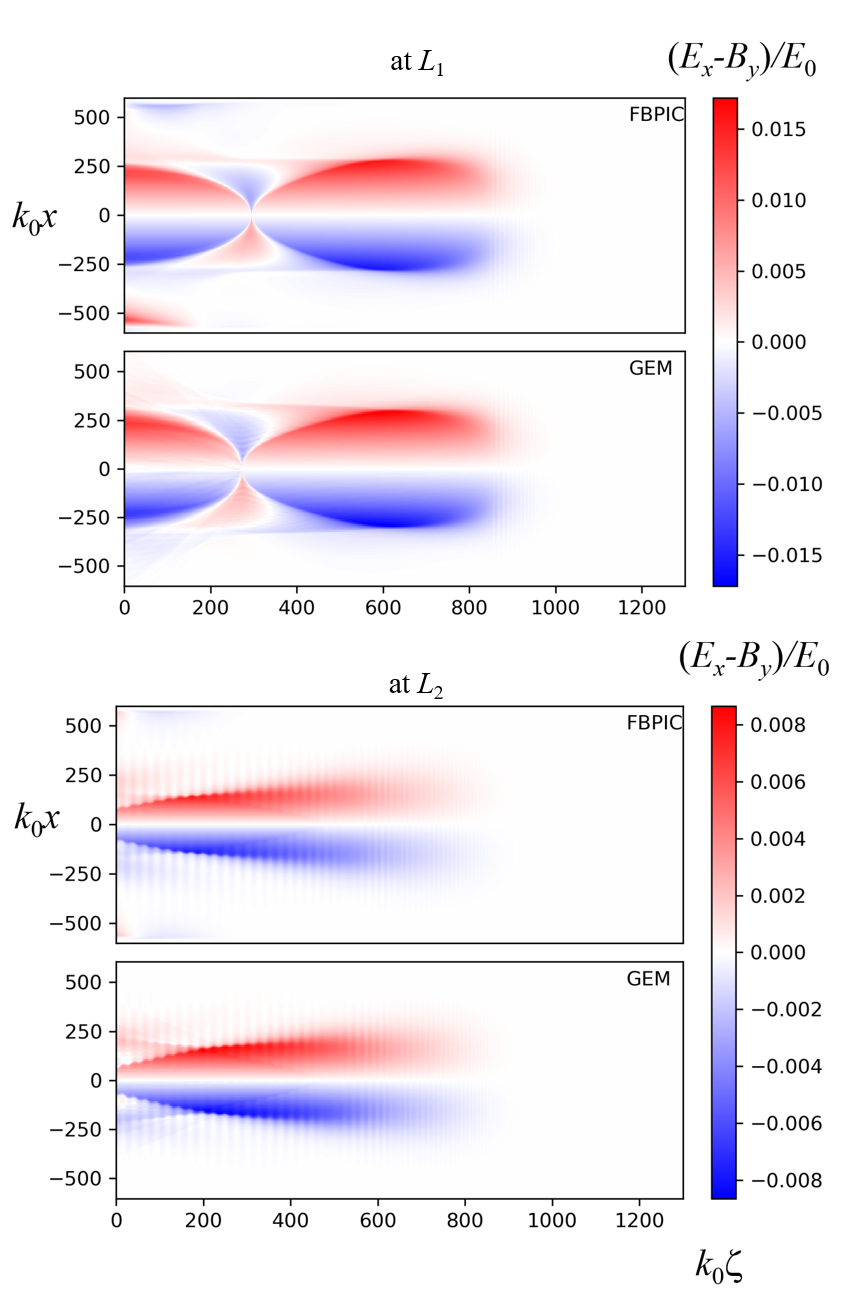}
\caption{2d distributions of the focusing $E_x-B_y$ obtained in FBPIC and GEM-PIC simulations.}
\label{Fig:Wake2Dfocus}
\end{figure}

The focusing field $E_x-B_y$  shown in Fig.~\ref{Fig:Wake2Dfocus} gives the focusing field of the wakefield generated by the laser at the positions $L_1$ and $L_2$ obtained in both codes.  

\begin{figure}
\includegraphics[width=10cm]{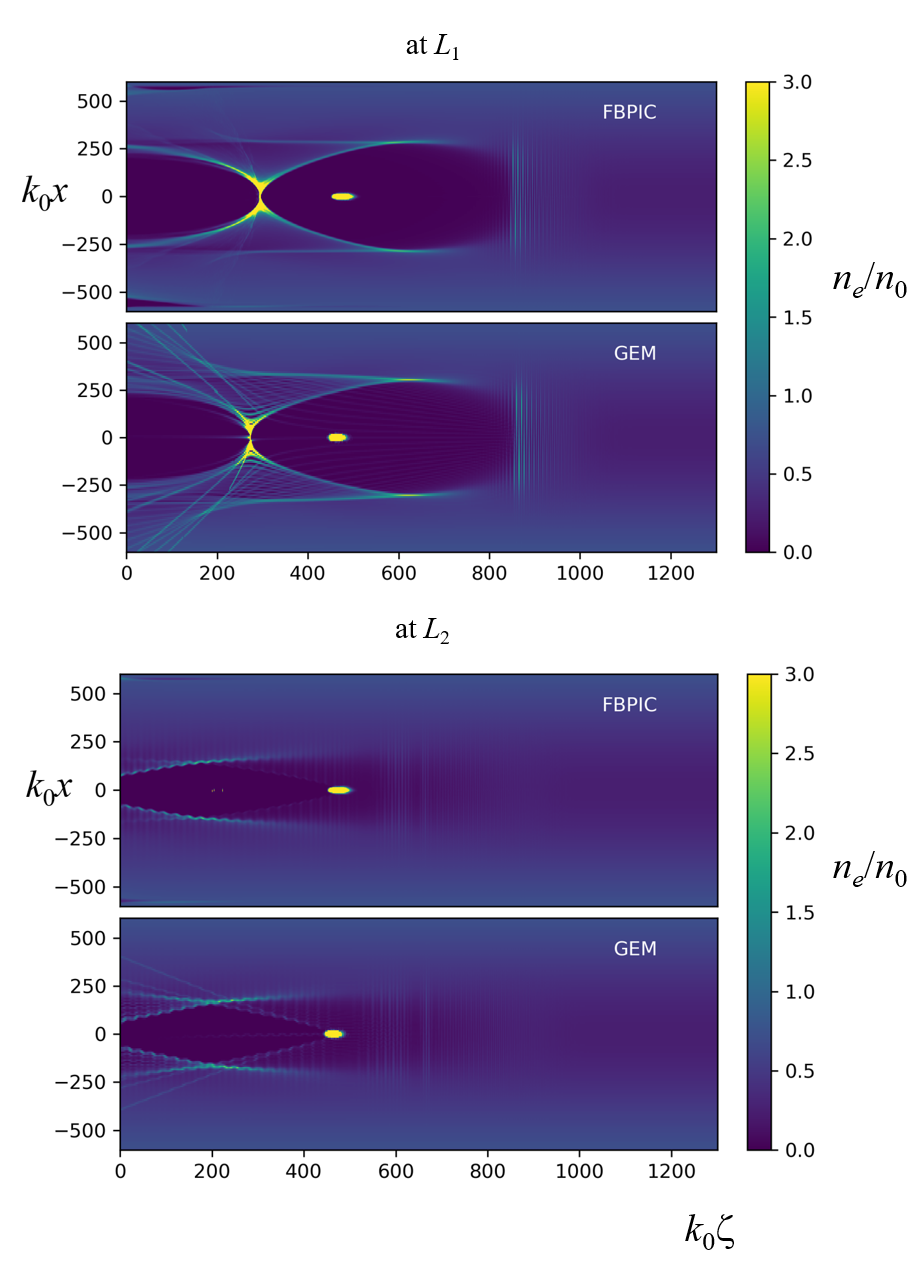}
\caption{2d density distributions of electrons. The trapped electron bunch is well seen within the bubble.}
\label{Fig:ne2d}
\end{figure}

\begin{figure}
\includegraphics[width=10cm]{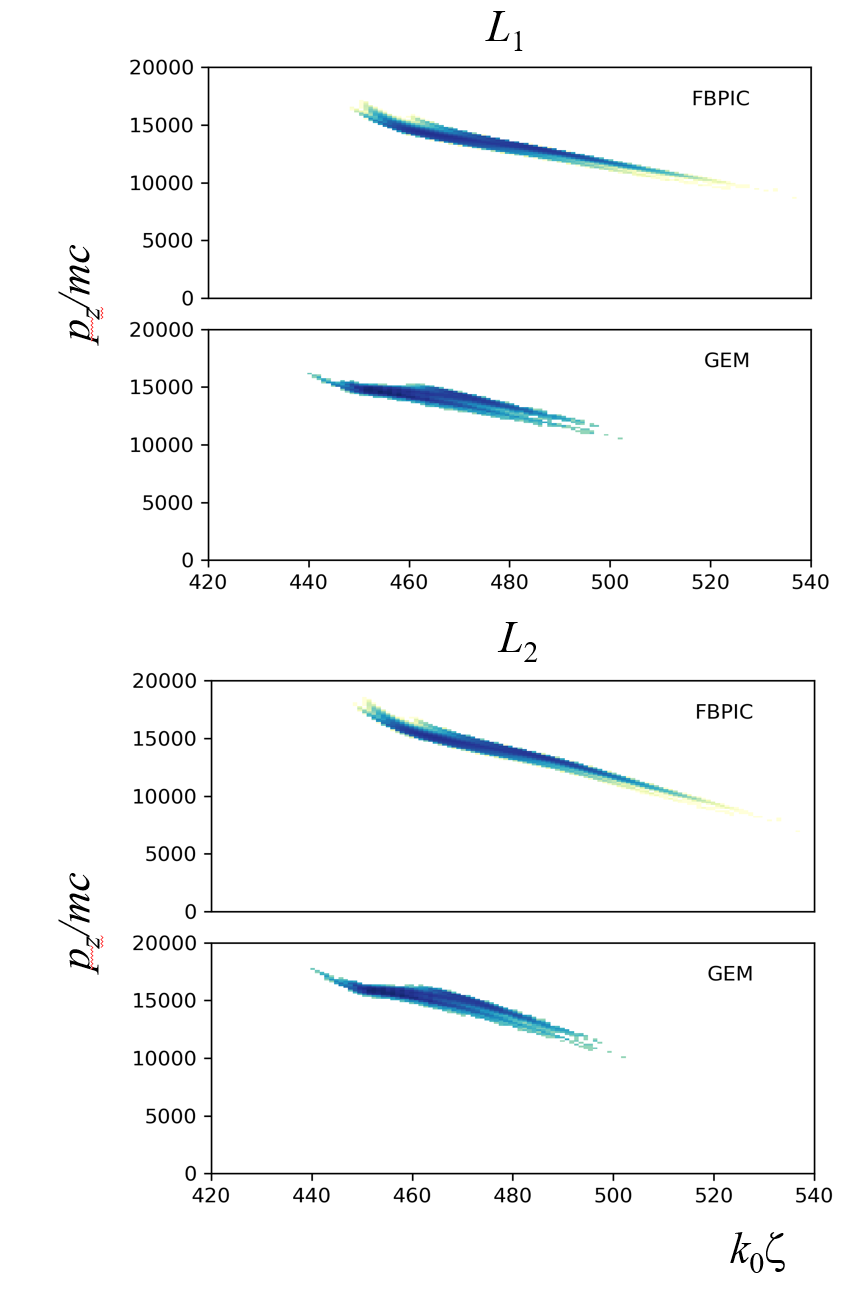}
\caption{Longitudinal phase space of the trapped electron bunch.}
\label{Fig:phasespace}
\end{figure}

Fig.~\ref{Fig:ne2d} compares the two-dimensional distribution of the electron density. The bubble shape, the witness bunch position and density agree well in both simulations.
Fig.~\ref{Fig:phasespace} presents the two-dimensional distribution of electron density.

The witness bunch parameters obtained in both codes are summarized in Table~\ref{TabL1}. Here we show the total trapped charge $Q$, the mean bunch energy $\mathcal{E}_0$, the r.m.s. energy spread  $\Delta \mathcal{E}$, the r.m.s. spatial dimensions $\sigma_x, ~\sigma_y,~\sigma_z$, the normalized emittances $\varepsilon_x$ and $\varepsilon_y$.

\begin{table}[h]
\centering
\caption{Witness bunch parameters at $s=L_1$ and $s=L_2$}
\begin{tabular}{|c|cc|cc|}
 \hline
 & $s=L_1$ & & $s=L_2$ & \\
 & FBPIC & GEM-PIC & FBPIC & GEM-PIC\\
 \hline
$Q$              & 67.61~pC & 92.61~pC & 67.61~pC & 92.61~pC \\
$\mathcal{E}_0$  & 6.95~GeV & 7.32~GeV& 7.32~GeV & 7.85~GeV\\
$\Delta \mathcal{E}$ & 6.63\% & 3.54\% & 6.91\% & 4.19\% \\
$\sigma_x$ & 7.6~$\mu$m & 6.42~$\mu$m & 7.34~$\mu$m & 6.39~$\mu$m \\
$\sigma_y$ & 4.07~$\mu$m & 4.95~$\mu$m & 4.31~$\mu$m & 4.96~$\mu$m \\
$\sigma_z$ & 10.72~$\mu$m & 8.37~$\mu$m & 10.67~$\mu$m & 8.34~$\mu$m \\
$\varepsilon_x$ & 5.14~$\mu$m & 4.02~$\mu$m & 5.11~$\mu$m & 4.16~$\mu$m \\
$\varepsilon_y$ & 1.71~$\mu$m & 2.41~$\mu$m & 1.77~$\mu$m & 2.48~$\mu$m \\
\hline
\end{tabular}
\label{TabL1}
\end{table}

Given that the total laser propagation length in the simulation was $L_2=2.5\cdot 10^6\,k_0^{-1}$, the overall agreement between both codes can be stated as reasonable.

\section{Conclusion and Outlook}
We have developed an efficient Galilean-boosted electromagnetic PIC algorithm that allows for very efficient simulations of laser-driven wakefield acceleration in plasmas. This algorithm closes the gap between the small scale defined by the laser wavelength (in micrometer range) and the huge acceleration length ranging from centimeters to meters. The slow coordinate grid step is limited by the characteristic evolution distance of the driver and not by the wavelength. This may provide acceleration in computational resources by many orders of magnitude as compared to the general electromagnetic PIC codes.

The scheme above is defined within a single step along the $s-$coordinate.
Thus, the grid step $\Delta$ in this direction can vary arbitrarily
at every level. There is no limitation at all on the uniformity of steps along
$s$. Integration along the $\tau-$coordinate uses the leap-frog method. This means that step $h$ cannot be abruptly changed from one cell to another without loss of accuracy. However, a smooth
change of the steps inside the box is possible without compromising the second-order accuracy too much. This change of stepping can be
very large, up to several orders of magnitude. It is important that the adjacent
grid steps do not differ significantly. Rather, the transition must
remain smooth. This can allow for resolution even of the x-ray wavelength locally,
if required.

\section*{Acknowledgements}

This work was supported by DFG and BMBF. The authors gratefully acknowledge the Gauss Centre for Supercomputing e.V. (gauss-centre.eu) for funding this project (lpqed) by providing computing time through the John von Neumann Institute for Computing (NIC) on the GCS Supercomputer JUPITER at Julich Supercomputing Centre (JSC).

\appendix

\section{Dispersion relation in 3D}
\label{DispApp}

We assume a plane wave
\[
{\bf E}=\hat{{\bf E}}\exp\left(-i\omega s+i\kappa \zeta+i{\bf k}_\perp\cdot{\bf r}_\perp\right),
\]
and model the plasma response using the cold-fluid equation
\[
-\partial_{\zeta}{\bf j}=\omega_{p}^{2}{\bf E},
\]
which in Fourier space gives
\[
{\bf j}= i\frac{\omega_p^2}{\kappa}{\bf E}.
\]

\subsection{Vacuum dispersion}

In vacuum ($\omega_p=0$) the analytical dispersion relation reads
\[
\omega^{2}= k_{\parallel}^{2}+k_{\perp}^{2}.
\]
Introducing the Galilean-shifted frequency
\[
\Omega=\omega-k_{\parallel},
\]
we obtain
\[
\Omega=\sqrt{k_{\parallel}^{2}+k_{\perp}^{2}}-k_{\parallel}
=\frac{k_{\perp}^{2}}{\sqrt{k_{\parallel}^{2}+k_{\perp}^{2}}+k_{\parallel}}.
\]
\replaced[id=R2]{In the paraxial limit $k_{\parallel}^{2}\gg k_{\perp}^{2}$ this reduces to}{Setting $c = 1$ and taking the limit $k_\parallel \gg k^2_\perp$ yields  }
\[
\Omega\simeq\frac{k_{\perp}^{2}}{2k_{\parallel}}
\left(1-\frac{k_{\perp}^{2}}{4k_{\parallel}^{2}}+\ldots\right).
\]
\subsection{TM polarization and analytical dispersion}

We now consider the TM polarization in the $(y,z)$ geometry, assuming that
\[
k_x=0,\qquad E_y=B_x=j_y=j_z=0.
\]
Under these assumptions, the reduced Maxwell system becomes
\begin{align}
\partial_s (E_x + B_y) &= i k_y B_z - J_x, \label{tm1}\\
2\partial_\zeta (E_x - B_y) &= - i k_y B_z + J_x, \label{tm2}\\
\partial_\zeta B_z &= - i k_y E_x, \label{tm3}\\
-\partial_\zeta J_x &= \omega_p^2 E_x. \label{tm4}
\end{align}

Using a Fourier ansatz $\propto \exp(-i\omega s+i\kappa\zeta)$, Eqs.~(\ref{tm3})--(\ref{tm4}) give
\[
B_z=-\frac{k_y}{\kappa}E_x,\qquad
J_x=i\frac{\omega_p^2}{\kappa}E_x.
\]
Substituting into Eq.~(\ref{tm1}) yields
\[
(-i\omega+i\kappa)E_x
=-\frac{i}{2\kappa}(k_y^2+\omega_p^2)E_x,
\]
which leads to the analytical dispersion relation
\begin{equation}
\boxed{
\omega=\kappa+\frac{k_y^2+\omega_p^2}{2\kappa}
}
\label{adisp2d}
\end{equation}
or, equivalently,
\[
\Omega\equiv\omega-\kappa=\frac{k_y^2+\omega_p^2}{2\kappa}.
\]

\subsection{Numerical dispersion}

We now derive the numerical dispersion relation. The 2D scheme extends the 1D case~(\ref{num1d}) and reads
\begin{align}
\frac{R^k_{n}-R^{k-1}_n}{\Delta } &=
-\frac{1}{4}\left(J^k_{n-1/2}+J^{k-1}_{n-1/2}
+J^k_{n+1/2}+J^{k-1}_{n+1/2}\right), \\
-2\frac{L^k_{n+1}-L^k_n}{h} &= -J^k_{n+1/2}-ik_y \hat{B}_z, \\
\frac{(B_z)^k_{n+1}-(B_z)^k_n}{h} &= -ik_y \hat{E}_x, \\
-\frac{J^k_{n+1/2}-J^k_{n-1/2}}{h} &= \omega^2_p E^k_n.
\label{num2d}
\end{align}

Using a discrete Fourier ansatz, the linear system can be written as

\[
\begin{bmatrix}
-\frac{2i}{\Delta }\tan(\omega\Delta/2) &
-\frac{2i}{\Delta }\tan(\omega\Delta/2) &
ik_y & \cos(\kappa h/2) \\
-\frac{4i}{h}\sin(\kappa h/2) &
+\frac{4i}{h}\sin(\kappa h/2) &
ik_y & 1 \\
ik_y & 0 & -\frac{2i}{\Delta \zeta} \sin(\kappa h/2)  & 0 \\
-\omega_p^2 & 0 & 0 &
-\frac{2i}{h}\sin(\kappa h/2)
\end{bmatrix}
\begin{bmatrix}
E_x \\ B_y \\ B_z \\ J_x
\end{bmatrix}
=0.
\]

Setting the determinant to zero yields the numerical dispersion relation
\begin{align}
\tan\left(\frac{\omega \Delta }{2}\right)=
-\frac{
2\Delta h (2k_y^2\sin(\kappa h/2)+\omega_p^2\sin(\kappa h))}
{h^2 (k_y^2+\omega_p^2) + 8\left(\cos(\kappa h)-1\right)}.
\label{scheme2d}
\end{align}
In the continuous limit $\Delta , h\to0$ this relation reduces to the analytical result~(\ref{adisp2d}).

\section{Analytical solution}
\label{DispAppSolution}

To solve system (\ref{RxN})-(\ref{eqall}), it is key to find updated longitudinal fields $E_{z0}^1$  and $B_{z0}^1$. The analytical solution is given below. 

First, we define

\begin{eqnarray}
K^2 &=& k_x^2 + k_y^2, \nonumber \\
D &= &16 + h(2\Delta+h)K^2, \nonumber 
\end{eqnarray}

\begin{eqnarray}
J_x &= &j_{x-1/2}^0+j_{x+1/2}^0+j_{x-1/2}^1+j_{x+1/2}^1, \nonumber \\
J_y &=& j_{y-1/2}^0+j_{y+1/2}^0+j_{y-1/2}^1+j_{y+1/2}^1, \nonumber 
\end{eqnarray}

\begin{eqnarray}
W_x &= &-2L_{x1}^1+R_{x0}^0+R_{x1}^1, \nonumber \\
W_y &= &-2L_{y1}^1+R_{y0}^0+R_{y1}^1.\nonumber 
\end{eqnarray}
The numerator terms can then be written as
\begin{eqnarray}
T_1 &= &E_{z1}^1(16 - h^2K^2), \nonumber  \\
T_2 &= &-16h\,j_{z+1/2}^1, \nonumber \\
T_3 &= &-i h \Delta (k_x J_x + k_y J_y), \nonumber  \\
T_4 &= &2 h \Delta\,E_{z0}^0K^2, \nonumber \\
T_5 &= &2 i h^2 (k_x j_{x+1/2}^1 + k_y j_{y+1/2}^1), \nonumber \\
T_6 &= &4 i h (k_x W_x + k_y W_y). \nonumber 
\end{eqnarray}
The updated longitudinal electric field becomes
\begin{equation}
E_{z0}^1=\frac{T_1 + T_2 + T_3 + T_4 + T_5 + T_6}{D}. \label{eq:EzSolved}
\end{equation}
We repeat the same procedure for the longitudinal magnetic field:
\begin{eqnarray}
U_x &=& 2L_{x1}^1+R_{x0}^0+R_{x1}^1, \nonumber \\
U_y &= &2L_{y1}^1+R_{y0}^0+R_{y1}^1 \nonumber 
\end{eqnarray}
with the numerator terms 
\begin{eqnarray}
Q_1 &=& B_{z1}^1(16 - h^2K^2), \nonumber \\
Q_2 &= &2 i h^2\,(k_x j_{y+1/2}^1 - k_y j_{x+1/2}^1), \nonumber \\
Q_3 &= &i h \Delta (k_x J_y - k_y J_x), \nonumber \\
Q_4 &= &-2 h \Delta\,B_{z0}^0K^2, \nonumber \\
Q_5 &= &4 i h (k_y U_x -  k_x U_y) \nonumber 
\end{eqnarray}
and obtain
\begin{equation}
B_{z0}^1=\frac{Q_1 + Q_2 + Q_3 + Q_4 + Q_5}{D} \label{eq:BzSolved}
\end{equation}
As soon as we know the updated longitudinal fields $E_{z0}^1$  and $B_{z0}^1$ (\ref{eq:EzSolved})-(\ref{eq:BzSolved}), the transverse field combinations $R_{x0}^1$, $R_{y0}^1$, $L_{x0}^1$ and $L_{y0}^1$ and thus the full set of updated fields ${\bf E}_0^1,\,{\bf B}_0^1,$ can be directly expressed from Eqs. (\ref{RxN})-(\ref{LyN}).

\bibliographystyle{unsrt}
\bibliography{sample.bib}

\end{document}